\documentclass[aps,prl,twocolumn]{revtex4}
\usepackage{amsmath}
\usepackage{graphicx}
\usepackage{dcolumn}
\usepackage{longtable}

\newcommand{\be}{\begin{equation}}
\newcommand{\ee}{\end{equation}}
\newcommand{\bea}{\begin{eqnarray}}
\newcommand{\eea}{\end{eqnarray}}
\newcommand{\bse}{\begin{subequations}}
\newcommand{\ese}{\end{subequations}}

\begin{document}

\title{The hydrogen atom in $D=3-2\epsilon$ dimensions}

\author{Gregory S. Adkins}
\email[]{gadkins@fandm.edu}
\affiliation{Franklin \& Marshall College, Lancaster, Pennsylvania 17604}

\date{\today}

\begin{abstract}
The nonrelativistic hydrogen atom in $D=3-2\epsilon$ dimensions is the reference system for perturbative schemes used in dimensionally regularized nonrelativistic effective field theories to describe hydrogen-like atoms.  Solutions to the $D$-dimensional Schr\"odinger-Coulomb equation are given in the form of a double power series.  Energies and normalization integrals are obtained numerically and also  perturbatively in terms of $\epsilon$.  The utility of the series expansion is demonstrated by the calculation of the divergent expectation value $\langle (V')^2 \rangle$.
\end{abstract}


\maketitle


For over a century the hydrogen atom has been a touchstone of fundamental physics.  From the earliest days of quantum physics, the challenge to understand the structure and behavior of hydrogen has been a driver of new developments and has stimulated the craft of countless innovators in the field. \cite{Rigden82,Rigden02}  One can mention the Bohr Model, Sommerfeld's relativistic hydrogen atom, wave and matrix mechanics at the birth of quantum mechanics, the development of QED to explain the Lamb Shift and the electron's anomalous moment as observed in hydrogen, detection of the hydrogen Bose-Einstein condensate, current experiments on antihydrogen, and its role in the ``proton size puzzle'' as developments in which hydrogen has played a central role.  Hydrogen is also a model for a number of ``exotic atoms'' including positronium, muonium, muonic hydrogen, hydrogen-like ions, charmonium, and bottomonium.  Deep understanding of hydrogen and hydrogen-like systems has been and continues to be crucially important.

Much of the modern work on hydrogen and its exotic siblings is based on the effective quantum field theories NRQED and NRQCD (non-relativistic quantum electro- and chromo-dynamics) \cite{Caswell86,Bodwin95,Hoang02,Petrov15}.  These theories build up the dynamics of both electro- or chromo-interactions as well as relativity as perturbations on a non-relativistic base.  Most of the modern work in NRQED/QCD uses dimensional regularization to control both ultraviolet and infrared divergences.  Consequently, the notion of a non-relativistic hydrogen-like system in $D=3-2\epsilon$ dimensions often plays a role as the lowest order of a perturbative development.  A typical calculation might involve the energy $E$ and the value of the wave function at contact $\psi(r=0)$ of this $D$-dimensional system \cite{Czarnecki99,Jentschura05}.  Usually, all divergences are arranged to cancel before values of the energy levels or wave function are actually needed, in which case the $D \rightarrow 3$ limit can be taken first, and traditional results for $E$ and $\psi(r=0)$ employed.  More generally, for higher-order calculations it is crucial to have as much flexibility in calculational approach as possible, and the restriction to schemes where all divergences must cancel first before making use of explicit forms for the wave functions is too limiting.  A detailed understanding of the system that is being used as a basis for perturbation theory is required.

In this work I describe the solution to the $D$-dimensional Schr\"odinger-Coulomb equation for non-integral $D$ near $D=3$.  Separation of variables is used to write the solution as a radial function times an angular function describing the orbital angular momentum.  The radial solution can be expressed as a power series of a novel type.  With the help of this series solution to handle the small-distance regime, the radial equation can be solved numerically using standard techniques.  I also work out the perturbative expansions for the energy and wave function at contact in terms of the small parameter $\epsilon=(3-D)/2$.  For energies, results through $O(\epsilon^2)$ are obtained with estimates for the $O(\epsilon^3)$ terms, while for the wave functions at contact results through $O(\epsilon)$ are obtained with estimates for the $O(\epsilon^2)$ terms.

Much of the earlier work on hydrogen in $D$ dimensions made use of a strict $1/r$ potential instead of the physical $1/r^{D-2}$ potential implied by Gauss' Law, or was restricted to an integral number of dimensions.  Quantum mechanics with the physical potential in non-integral dimension has been studied by Andrew and Supplee \cite{Andrew90}, Morales \cite{Morales96}, and is reviewed in \cite{Dong11} with additional references.

The $D$-dimensional Schr\"odinger-Coulomb equation is
\be \label{SCED_1}
-\frac{1}{2m} \vec \nabla^2 \bar \psi(\vec x\,) + \bar V(r) \bar \psi(\vec x\,) = \bar E \bar \psi(\vec x\,) ,
\ee
where we use the bar to signify $D$-dimensional quantities, with no bar for 3-dimensional ones.  The potential energy
\be \label{coulomb_potential_bar}
\bar V(r) = - \frac{\Gamma \left (D/2-1 \right ) {\bar \mu}^{2 \epsilon} Z \alpha}{\pi^{D/2-1} r^{D-2}}
\ee
arises as the $D$-dimensional Fourier transform of the momentum-space Coulomb interaction term $-4 \pi Z \alpha \bar \mu^{2 \epsilon}/p^2$.  It follows that (\ref{SCED_1}) is the lowest approximation for the study of hydrogen-like atoms in NRQED/QCD (where $\hbar=1$, $\alpha$ is the fine structure constant, and $Z$ is the nuclear charge in units of the electron charge magnitude).  The potential $\bar V(r)$ can also be deduced from the requirement that the electric field derived from it satisfy Gauss's law in $D$ dimensions, or equivalently that the potential satisfy the $D$-dimensional Poisson equation with a point charge source.  The mass scale $\mu$ has been introduced to ensure that $Z \alpha$ remains dimensionless in $D$ dimensions, and $\bar \mu$ (with $\bar \mu^2 \equiv \mu^2 e^{\gamma_E}/[4 \pi]$) is the corresponding $\overline{\text{MS}}$ scale.   It is convenient to separate variables in the Schr\"odinger equation using spherical coordinates.  The $D$-dimensional Laplacian can be written as
\be
\vec \nabla^2 = \partial_r^2 + \frac{D-1}{r} \partial_r - \frac{L^2}{r^2}
\ee
where $r=\big ( \sum_i x_i^2 \big )^{1/2}$ is the usual radius and $L^2=\sum_{i < j} L_{i j}^2$ with $L_{i j}=-i (x_i \partial_j - x_j \partial_i)$ is the angular momentum squared.  We separate variables in the wave function according to $\bar \psi(\vec x\,) = \bar R(r) Y(\hat x)$ where $\hat x = \vec x /r$.  The angular functions $Y(\hat x)$ are eigenstates of $L^2$ \cite{Louck60,Avery89,Avery10}:
\be
L^2 Y_{\ell}(\hat x) = \ell (\ell+D-2) Y_{\ell}(\hat x),
\ee
where the allowed quantum numbers $\ell$ are 0, 1, 2, $\cdots$, just as in an integral numbers of dimensions.  The $D$-dimensional angular functions have $(2\ell+D-2)(\ell+D-3)!/\left ( \ell! [D-2]!\right )$ independent components.  An explicit representation is given by the symmetric traceless harmonic polynomials.  For example, the lowest few are $Y_0(\hat x) = A_0$, $Y_{1 i}(\hat x) = A_1 \hat x_i$, $Y_{2 i j}(\hat x) = A_2 \left ( \hat x_i \hat x_j - \delta_{i j}/D \right )$, containing $1$, $D$, and $(D-1)(D+2)/2$ independent components, where the $A_\ell$ are appropriate normalization factors.

The radial equation is the object of our main concern.  It is
\bea \label{radial_SCE_0}
&\hbox{}& \hskip -0.5cm \frac{1}{2 m} \left \{ - \partial_r^2 - \frac{D-1}{r} \partial_r + \frac{\ell (\ell+D-2)}{r^2} \right \} \bar R_{n \ell}(r) \nonumber \\
&\hbox{}& \quad \quad + \bar V(r) \bar R_{n \ell}(r) = \bar E_{n \ell} \bar R_{n \ell}(r) .
\eea
We follow the usual steps of first working out the leading short and long distance behavior of $\bar R$.  The $r \rightarrow 0$ limit of (\ref{radial_SCE_0}) shows that $\bar R(r) \rightarrow r^\ell$ for small $r$.  We also find $\bar R(r) \rightarrow e^{-\bar \gamma r}$ for large $r$ where $\bar E = - \bar \gamma^2/(2m)$.  We define a new function $L_{n \ell}(\rho)$ according to
\be \label{def_Lnell}
\bar R_{n \ell}(r) = \bar \phi_{n \ell} \Omega_{D-1}^{1/2} \left ( \frac{(n+\ell)!}{n (n-\ell-1)!} \right )^{1/2} \frac{\rho^\ell e^{-\rho/2}}{(2\ell+1)!} L_{n \ell}(\rho)
\ee
where $\rho \equiv 2 \bar \gamma r$ is dimensionless and $\Omega_n \equiv 2 \pi^{\frac{n+1}{2}}/\Gamma(\frac{n+1}{2})$ is the surface area of a unit $n$-sphere.  We normalize $L_{n \ell}(r)$ so that $L_{n \ell}(0)=1$.  We see that $\bar \phi_{n 0} = \underset{r \rightarrow 0}{\lim} \bar \psi_{n 0}(r)$ is the S-state wave function at the origin (at ``contact''), and generally $\bar \phi_{n \ell}$ is proportional to $\underset{r \rightarrow 0}{\lim} \bar R_{n \ell}(r)/r^\ell$.  When expressed in terms of $\rho$ and $L_{n \ell}(\rho)$, the radial equation becomes
\bea \label{radial_SCE_2}
&\hbox{}& \bigg \{ \partial_\rho^2 + \left ( \frac{2 (\ell+1-\epsilon)}{\rho} - 1 \right ) \partial_\rho - \frac{\ell+1-\epsilon}{\rho} \nonumber \\
&\hbox{}& \hskip 2.0cm  + \frac{\bar{n} \rho^{2 \epsilon}}{\rho} \bigg \} L_{n \ell}(\rho) = 0 
\eea
where
\be \label{formula_for_nbar}
\bar{n} \equiv \frac{m Z \alpha}{\bar \gamma} \frac{\Gamma(1/2-\epsilon)}{\pi^{1/2-\epsilon}} \left ( \frac{\bar \mu}{2 \bar \gamma} \right )^{2 \epsilon} .
\ee
In three dimensions, $\bar n$ would be the principal quantum number $n$ and $\bar \gamma$ would be the momentum scale factor $m Z \alpha/n$.

We intend to find a series solution for (\ref{radial_SCE_2}) about the origin.  Since $\rho^{2\epsilon-1}$ is not analytic in a region containing the origin for most values of $\epsilon$, the usual type of series solution won't work.  We require the more general form
\be \label{series_soln}
L_{n \ell}(\rho) = \sum_{j=0}^\infty \sum_{k=0}^j a_{j k} \bar{n}^k \rho^{j+2 \epsilon k} .
\ee
Using (\ref{series_soln}) in (\ref{radial_SCE_2}) and assuming that all powers $\rho^{j+2 \epsilon k}$ are independent, we obtain the recursion relation:
\be \label{recursion}
a_{j k} = \frac{a_{j-1, k} \left ( j+\ell+\epsilon [ 2k-1 ] \right ) - a_{j-1, k-1}}{(j+2 \epsilon k) \left ( j+2 \ell+1+2 \epsilon [k-1] \right )} .
\ee
Using (\ref{recursion}) and the initial condition $a_{0 0}=1$, it is easy to calculate as many coefficients $a_{j k}$ as desired and obtain a convergent series solution near $\rho = 0$.  (When $\epsilon \rightarrow 0$, the solution for $a_{j k}$ is $(-1)^j s_{-(\ell+1)}(j,k) / \big ( j! [2\ell+2]^{\bar j} \big )$, where the $s_a(j,k)$ are ``non-central Stirling numbers of the first kind'' as defined by Koutras \cite{Koutras82} and $n^{\bar j}$ is the rising factorial $n^{\bar j} = n (n+1) \cdots (n+j-1)$.  In this limit the $L_{n \ell}$ reduce to the usual associated Laguerre polynomials.)  We use the series to find $L_{n \ell}(\rho)$ in a small region ($0 \le \rho \le \rho_0$) around the origin and extend that region to $0 \le \rho < \infty$ using standard numerical methods to solve (\ref{radial_SCE_2}).  We developed a procedure to home in on acceptable values of $\bar n$ for which $L_{n \ell}(\rho)$ can be normalized (as in the integral (\ref{normalization_int_1}) below).  For each value of $\ell$ we labeled these solutions by the ``radial quantum number'' $n_r$ taking values 0, 1, 2, $\cdots$.  We also define the standard principal quantum number $n$ with $n_r=n-\ell-1$, which takes positive integer values starting with $\ell+1$ for each value of $\ell$.  The acceptable values of $\bar n$ with $\epsilon=0.001$ for the low-lying states are shown in Table~\ref{results_table} as $\bar n^{\text{DE}}$.  We used the numerical solutions to compute values for the integrals
\be \label{normalization_int_1}
I_{n \ell} \equiv \frac{(n+\ell)!}{2n n_r! [(2\ell+1)!]^2} \int_0^\infty d \rho \, \rho^{D-1+2\ell} e^{-\rho} \Bigl [ L_{n \ell}(\rho) \Bigr ]^2
\ee
that are related to the normalization of the corresponding states.  These appear in the table as $I^\text{DE}$.  Were $D=3$, the $I_{n \ell}$ integrals would all be one.
\begin{table*}
\begin{center}
\caption{\label{results_table}  Numerical values for several quantities for low-lying states: the second-order perturbation theory matrix element $\kappa$; $\xi^{[2]}$, the $O(\epsilon^2)$ part of $\xi \equiv \bar n/n$; the perturbation result for $\bar n=n \xi$ when $\epsilon=0.001$ (through $O(\epsilon^2)$); the value of $\bar n$ when $\epsilon=0.001$ found directly from the differential equation; an estimate for $\xi^{[3]}$, the $O(\epsilon^3)$ coefficient of $\xi$; the perturbation result for the normalization integral $I$ when $\epsilon=0.001$ (through $O(\epsilon)$); the normalization integral $I$ when $\epsilon=0.001$ found by numerical integration after solving the differential equation; and an estimate for $I^{[2]}$, the $O(\epsilon^2)$ coefficient of $I$.  The uncertainties in the values for $\bar n^\text{DE}$ and $I^\text{DE}$ are no more than one in the least significant digit.}
\begin{ruledtabular}
\begin{tabular}{cccccccccc}
$n$ & $\ell$ & $\kappa_{n \ell}$ & $\xi^{[2]}$ & $\bar n^\text{pert. \!\!th.}$ & $\bar n^\text{DE}$ & $\xi^{[3]}$ & $I^\text{pert. \!\!th.}$ & $I^\text{DE}$ & $I^{[2]}$ \\
\hline\noalign{\smallskip}
\noalign{\smallskip} 1 & 0 & 0.447424 & 0.264439 & 0.998154696 & 0.998154698 & 2.4 &1.006734 & 1.006747 & 13 \\
\hline
\noalign{\smallskip} 2 & 0 & 0.024322 & -0.621005 & 1.995307621 & 1.995307635 & 7.3 & 1.010814 & 1.010857 & 43 \\
\noalign{\smallskip} 2 & 1 & 0.125789 & 2.859042 & 1.993981247 & 1.993981249 & 0.9 & 1.012147 & 1.012206 & 58 \\
\hline
\noalign{\smallskip} 3 & 0 & -0.039708 & -0.228670 & 2.991462608 & 2.991462640 & 10.8 & 1.013227 & 1.013295 & 68 \\
\noalign{\smallskip} 3 & 1 & 0.342397 & 2.811021 & 2.989971727 & 2.989971739 & 4.0 & 1.016394 & 1.016507 & 113 \\
\noalign{\smallskip} 3 & 2 & 0.539319 & 5.240299 & 2.988779015 & 2.988779009 & -1.8 & 1.017994 & 1.018135 & 141 \\
\hline
\noalign{\smallskip} 4 & 0 & 0.050335 & 0.533015 & 3.986953191 & 3.986953242 & 12.9 & 1.014946 & 1.015034 & 88 \\
\noalign{\smallskip} 4 & 1 & 0.562100 & 3.210384 & 3.985363900 & 3.985363925 & 6.1 & 1.019657 & 1.019824 & 167 \\
\noalign{\smallskip} 4 & 2 & 0.839676 & 5.435611 & 3.984039468 & 3.984039469 & 0.2 & 1.022323 & 1.022547 & 223 \\
\noalign{\smallskip} 4 & 3 & 1.062267 & 7.338238 & 3.982904221 & 3.982904201 & -5.1 & 1.024038 & 1.024302 & 263 \\
\end{tabular}
\end{ruledtabular}
\end{center}
\end{table*}

As a complement to the numerical solutions obtained above we have also worked out results for $\bar n$ and $I$ using perturbation theory in the small parameter $\epsilon$.  This was done in order to confirm the consistency of the whole $D$-dimensional procedure and for use in the evaluation of coordinate-space matrix elements.  The zeroth-order problem for this perturbative calculation is also $D$-dimensional, but with a potential $\tilde V(r) = -Z \alpha/r$.  It is essential that the zeroth-order problem be $D$-dimensional, as the two Hamiltonians and the perturbation must be hermitian {\it in the same space\/}.  Fortunately this zeroth-order problem has an exact solution \cite{Alliluev57,Nieto79} as described by Nieto.  The radial equation in this case is identical to (\ref{radial_SCE_2}) except that the potential term $\bar n \rho^{2 \epsilon}/\rho$ is replaced by $\tilde n/\rho$.  The exact solution to this zeroth-order problem can be expressed as
\be \label{Ddim_radial_wf}
\tilde R_{n \ell}(r) = {\cal N}(n,\ell) e^{-\rho/2} \rho^\ell L_{n-\ell-1}^{2\ell+1-2\epsilon}(\rho)
\ee
where $\rho = 2 \tilde \gamma r$, $\tilde n = n-\epsilon$, $\tilde \gamma = m Z \alpha/\tilde n$.  The bound state energy is $\tilde E_n = -\tilde \gamma^2/(2m)$, and the normalization constant is given in \cite{Nieto79}.  The associated Laguerre polynomials are defined in the standard way: $L_n^\alpha(x) = \sum_{j=0}^n \binom{n+\alpha}{n-j} \frac{(-x)^j}{j!}$.
The perturbation is
\be
H'=\bar V(r)-\tilde V(r)= -\frac{2 Z \alpha}{r} \Bigl ( \ln (\mu r) + \gamma_E \Bigr ) \epsilon + O(\epsilon^2) .
\ee
It is straightforward to work out the first energy correction:
\be
\Delta E_{n \ell}^{[1]} = 4 E_n \left ( L + H_{n+\ell} \right ) \epsilon ,
\ee
where $E_n=-\gamma^2/(2m)$ is the standard Bohr energy, $\gamma = m Z \alpha/n$, $L = \log ( \mu / [2 \gamma])$, and $H_n=\sum_{j=1}^n 1/j$ is the $n^\text{th}$ harmonic number.  My calculation of the second order energy correction makes use of the form for the reduced Schr\"odinger-Coulomb Green's function $\hat g_n$ given by Johnson and Hirschfelder \cite{Johnson79}.  I was not able to obtain a general formula for the $O(\epsilon^2)$ energy correction.  For any particular state I was able to obtain the $O(\epsilon^2)$ correction in terms of $\kappa_{n \ell}$ where
$E_n \kappa_{n \ell} = \big \langle V \ln(2 \gamma r) \hat g_n \ln(2 \gamma r) V \big \rangle$,
for which I could only obtain numerical results.  (In the calculation of $\kappa_{n \ell}$ it was adequate to use the $\epsilon \rightarrow 0$ limit of $H'$ and standard 3-dimensional expressions for the states and the reduced Green's function.)  For instance, the ground state energy has the expansion
\bea
\bar E_{1 0} &=& 1+ \epsilon \left \{ 4L+6 \right \} + \epsilon^2 \Bigl \{ 8L^2+16 L - 4 \gamma_E^2 \nonumber \\
&\hbox{}& + 15 -\zeta(2) + 4 \kappa_{1 0} \Bigr \} + O(\epsilon^3) ,
\eea
where $\gamma_E$ is the Euler-Mascheroni constant.  From the energies, we can obtain the series for $\bar \gamma = (-2 m \bar E)^{1/2}$, then $\bar n$ using (\ref{formula_for_nbar}), and finally $\xi = \bar n/n = 1 + \xi^{[1]} \epsilon + \xi^{[2]} \epsilon^2 + \xi^{[3]} \epsilon^3 + \cdots$.  The exact result for $\xi^{[1]}$ is $\xi^{[1]} = 2 \gamma_E - 2 H_{n+\ell} - 1/n$, and for the ground state one finds
\be
\xi_{1 0} = 1 + \epsilon \Bigl \{ 2 \gamma_E-3 \Bigr \} + \epsilon^2 \Bigl \{ 4\gamma_E^2 - 6 \gamma_E + 2 \zeta(2) - 2 \kappa_{1 0} \Bigr \} + O(\epsilon^3) .
\ee
Table~\ref{results_table} contains numerical results for $\xi^{[2]}$ as calculated using perturbation theory as well as estimates for $\xi^{[3]}$ obtained by a numerical exploration of the difference between $\bar n^\text{DE}$ and the truncated series $\bar n^\text{pert. \!\!th.} = n (1 + \xi^{[1]} \epsilon + \xi^{[2]} \epsilon^2)$ for various small values of $\epsilon$.  The series for $\xi$ seems well-behaved at least through $O(\epsilon^3)$.

Now we work out the perturbative result for $\bar \phi_{n \ell}$ describing the short-distance behavior of the wave function and the related result for the normalization integral (\ref{normalization_int_1}) for the radial functions $L_{n \ell}(\rho)$.  We can calculate $\bar \phi_{n \ell}$ from (\ref{def_Lnell}) as the short-distance limit
\be \label{def_phinell}
\bar \phi_{n \ell} = \Omega_{D-1}^{-1/2} \left ( \frac{n n_r!}{(n+\ell)!} \right )^{1/2} \frac{(2\ell+1)!}{(2\bar \gamma)^\ell} \lim_{r \rightarrow 0} \frac{1}{r^\ell} \bar R_{n \ell}(r) .
\ee
We use first-order perturbation theory based on the exact solution of the $D$--dimensional $1/r$ problem to find the $O(\epsilon)$ correction to the wave function and then to $\bar \phi_{n \ell}$.  Since the perturbation is purely radial, we can factor out the angular momentum dependence and write
\bea
\bar R_{n \ell}(r) &=& \tilde R_{n \ell}(r) + \int dr_1 \, r_1^{D-1} \hat{\tilde g}_{n \ell}(r,r_1) H'(r_1) \tilde R_{n \ell}(r_1) \nonumber \\
&+& O(\epsilon^2) ,
\eea
where $\hat{\tilde g}_{n \ell}(r,r_1)$ is the component of the reduced Green's function for the $D$--dimensional $1/r$ problem having angular momentum $\ell$.  The $O(\epsilon)$ correction here contains an explicit factor of $\epsilon$ in $H'$, so in order to get just the first order correction we can take $\epsilon \rightarrow 0$ in the rest and simply use the regular $3$--dimensional reduced Green's function and radial wave function.  The result for the expansion of $\bar \phi_{n \ell}$ is
\begin{eqnarray} \label{phi_total}
\bar \phi_{n \ell} &=& \left ( \frac{\gamma^D}{\pi} \right )^{1/2} \biggl \{ 1 + \epsilon \biggl [ 3 L + 2n \, \text{diH}_+(n+\ell,-n_r) \nonumber \\
&\hbox{}& \hskip -1.0cm - n \left ( H_{n+\ell}^2-H_{n+\ell}^{(2)} \right ) + n \left ( H_{n_r}^2 + H_{n_r}^{(2)} \right ) + 2 H_{n+\ell} + 2 H_{2\ell+1} \nonumber \\
&\hbox{}& + \frac{1}{2} ( \ln \pi - \gamma_E) -2 +\frac{2}{n} -2n \zeta(2) \biggr ] + O(\epsilon^2) \biggr \} .
\end{eqnarray}
Here $H^{(2)}_n = \sum_{j=1}^n 1/j^2$ is a generalized harmonic number.  We have found it convenient to define as well the ``diharmonic'' numbers
\be \label{diharmonic_defs}
\text{diH}_\pm(n,m) \equiv \sum_{i=1}^n \frac{H_{m\mp 1\pm i}}{i} = \sum_{i=1}^n \sum_{j=1}^{m \mp 1\pm i} \frac{1}{i j} ,
\ee
in terms of which it is possible to express any diharmonic sum (a sum of $1/(i j)$ for positive integers $i$, $j$) over a region of the $i$, $j$ lattice having a boundary that includes vertical, horizontal, and diagonal sides of angle $\pm 45^\circ$ only.  The normalization integral $I_{n \ell}$ is connected to $\bar \phi_{n \ell}$ because the radial wave function is normalized in $D$-dimensional space: $1 = \int_0^\infty dr \, r^{D-1} \bar R_{n \ell}^2(r) = \bar \phi_{n \ell}^2 \Omega_{D-1} 2 I_{n \ell} / (2 \bar \gamma)^D$, so the normalization integral $I_{n \ell}$ must have the value
\be
I_{n \ell} = \frac{2^{D-2} \Gamma(D/2)}{\pi^{D/2-1}} \left ( \frac{\bar \gamma^D}{\pi \bar \phi_{n \ell}^2} \right ) .
\ee
We use our earlier result for $\bar \gamma$ and (\ref{phi_total}) for $\bar \phi_{n \ell}$ to write
\bea \label{expansion_for_Inell}
I_{n \ell} &=&  1 + 2 \epsilon \biggl [ n \Bigl ( H_{n+\ell}^2 - H_{n+\ell}^{(2)} \Bigr ) - n \Bigl ( H_{n_r}^2 + H_{n_r}^{(2)} \Bigr ) \nonumber \\
&\hbox{}& \hskip 0.0cm - 2n \, \text{diH}_+(n+\ell,-n_r) + 2n \zeta(2) + H_{n+\ell} \nonumber \\
&\hbox{}& \hskip 0.0cm  - 2 H_{2\ell+1} + 1 - \frac{1}{2n} + \gamma_E \biggr ] + O(\epsilon^2) .
\eea
This result for $I_{n \ell}$ truncated at $O(\epsilon)$ is given in Table~\ref{results_table} as $I^\text{pert. \!\!th.}$.  By comparison with the numerical result $I^\text{DE}$ for various small values of $\epsilon$, an estimate for the $O(\epsilon^2)$ term $I^{[2]}$ was also obtained and displayed in the table.

In this work we can see that $\bar n$ and $I$ play the role of scale invariant quantities independent of $\mu$ since they are determined directly from the scale-free differential equation (\ref{radial_SCE_0}).  On the other hand, the energy $\bar E$, momentum scale $\bar \gamma$, and the short distance wave function factor $\bar \phi$ all depend on $\mu$, as can be seen directly from their definitions and from the logarithms present in their series expansions.

The numerical approach developed here gives precise results for all $D$ in the range $2<D<4$.  This range is bordered by $D=2$, where the potential becomes logarithmic and the spectrum takes a qualitatively different form \cite{Asturias85,Eveker90}; and by $D=4$, where the potential and centrifugal terms merge and stable solutions do not exist (as reviewed in \cite{Bures15}).  We can compare our numerical results for the ground state energy to previous numerical results by Andrew and Supplee \cite{Andrew90}, who obtained a numerical solution to the Schr\"odinger equation directly, Morales \cite{Morales96}, who used the ``shifted $1/d$ method'', and Waldstein \cite{Waldstein08}, who used the variational method with trial function $r^a e^{-b r}$.  The results are given in Table~\ref{ground_state_energies} using the same units $m=\hbar=1$ and $4 \pi Z \alpha \bar \mu^{2 \epsilon}/\Omega_{D-1} = 1$ as in \cite{Andrew90,Morales96}.
\begin{table}
\begin{center}
\caption{\label{ground_state_energies}  Numerical values for the ground state energy as given by Andrew and Supplee (``A\&S'') \cite{Andrew90}, Morales \cite{Morales96}, Waldstein \cite{Waldstein08}, and  $\bar E_{n 0}=-\bar \gamma_{n 0}^2/(2 m_r)$ in the present work.  The dimensionless units suggested in \cite{Andrew90} are used.  The uncertainties in the energies computed here are no more than one in the least significant digit.}
\begin{ruledtabular}
\begin{tabular}{ccccc}
$D$ & A\&S & Morales & Waldstein & this work \\
\hline\noalign{\smallskip}
\noalign{\smallskip} 2.4 & -2.1678 & -2.1786 & -2.1667 & -2.176589 \\
2.8 & -0.8110 & -0.8011 & -0.8004 & -0.801097 \\
3.0 & -0.5000 & -0.5000 & -0.5000 & -0.500000 \\
3.4 & -0.1501 & -0.1502 & -0.1489 & -0.150171 \\
3.8 & -0.0087 & -0.0089 & -0.0077 & -0.008741
\end{tabular}
\end{ruledtabular}
\end{center}
\end{table}

As an example of the utility of knowing the series expansion for the wave function, we will evaluate $\langle (V')^2 \rangle_{n 0}$, an expectation needed when working out energy corrections for hydrogenic systems at $O(m \alpha^6)$.  This expectation value is easy to evaluate for $\ell > 0$ but is divergent, containing a $1/\epsilon$, when $\ell=0$, in which case
\be \label{expec_V'2}
\left \langle (V')^2 \right \rangle_{n 0} = \int_0^\infty dr \, r^{D-1} \left [ V'(r) \right ]^2 \bar R^2_{n 0}(r) .
\ee
We write $\bar R_{n 0}(r) = \bar \phi_{n 0} \Omega^{1/2}_{D-1} e^{-\rho/2} L_{n 0}(\rho)$ and use the series expansion $L_{n 0}(\rho) = \widehat L_{n 0}(\rho) + \big [ L_{n 0}(\rho) - \widehat L_{n 0}(\rho) \big ]$ where $\widehat L_{n 0}(\rho) = 1 + \rho/2 - \bar n \rho^{1+2\epsilon}/(2[1+2\epsilon])$.  The only divergence comes from the part of (\ref{expec_V'2}) containing ${\widehat L}^{\, 2}_{n 0}(\rho)$---the rest is finite and is relatively easy to evaluate for all $n$.  We find
\bea
\left \langle (V')^2 \right \rangle_{n 0} &=&  \pi m (Z \alpha)^3 \bar \phi_{n 0}^2 \bar \mu^{2\epsilon} \nonumber \\
&\hbox{}& \hskip -1.2cm \times \left \{ -\frac{2}{\epsilon} - 8 L + 8H_n + \frac{4}{3 n^2} - \frac{4}{n} - \frac{16}{3} \right \} .
\eea
This result for $\langle (V')^2 \rangle_{n 0}$ can also be obtained by a momentum space calculation.  Other dimensionally regularized expectation values needed at $O(m \alpha^6)$ coming from second order perturbation theory, not easily accessible to momentum space calculations, can also be obtained by means of the series expansion for the wave function.  The detailed information about the short distance behavior of the wave function contained in the double series expansion (\ref{series_soln}) allows for calculations involving arbitrary values of $n$ to be achieved completely in dimensional regularization.


\begin{acknowledgments} 
The author acknowledges helpful conservations with I. Waldstein.  This work was supported by the National Science Foundation through Grant No. PHY-1707489.
\end{acknowledgments}


     


\end{document}